\documentstyle[aps,pra,epsfig,amstex,amssymb,fancyheadings]{revtex}
\begin{document}
\title{\bf Spectrum of Neutral Helium in Strong Magnetic Fields}
 
\author{Matthew D. Jones and Gerardo Ortiz}
 
\address{
Theoretical Division, Los Alamos National Laboratory, \\
Los Alamos, NM 87545}

\author{David M. Ceperley} 
\address{
Department of Physics and National Center for Supercomputing Applications, \\
University of Illinois at Urbana-Champaign \\
     1110 West Green Street, Urbana, Il 61801 \\
}
 
\date{\today}
\maketitle
 
\begin{abstract}

We present extensive and accurate calculations for the
excited state spectrum of spin-polarized neutral helium in a range 
of magnetic field strengths up to $10^{12}$ G.
Of considerable interest to models of magnetic white dwarf stellar atmospheres,
we also present results for the dipole strengths of the low lying
transitions among these states.
Our methods rely on a systematically saturated basis set approach to 
solving the Hartree--Fock
self-consistent field equations, combined with an ``exact'' stochastic
method to estimate the residual basis set truncation error
and electron correlation effects.
We also discuss the applicability of the adiabatic approximation to
strongly magnetized multi-electron atoms.

\end{abstract}



\section{Introduction}

The electronic structure of simple atomic systems in strong external
fields remains poorly understood, despite considerable theoretical effort.
These systems are of critical importance in certain stellar environments,
in which very large magnetic fields have been inferred\cite{landstreet}.
A detailed knowledge of the spectra of light atoms (presumed to dominate
the atmospheres of compact stellar remnants) subjected to intense
magnetic fields would enable both observers and theorists to better refine 
their understanding of these astrophysical objects.
Unfortunately, only the spectrum of hydrogen has been adequately treated
thus far, by Rosner {\it et al.} in 1984\cite{rosner}.
This detailed work on hydrogen has been successfully applied
to the observed spectra\cite{ruder} from many magnetic white dwarf stars,
but several stars remain in which the spectra can not be accounted
for by hydrogen\cite{schmidt}, and in which the determination of the
strength and configuration of the stellar magnetic field would be greatly
aided by precise calculations of the spectrum of the next lightest element,
neutral helium.

The difficulty in theoretically treating atoms in strong magnetic fields 
lies in
the fact that magnetic and Coulomb forces are of nearly equal importance;
neither can be treated as a perturbation of the other.
In the uniform magnetic fields that we consider in this work (assumed to lie
in the $z$-direction), this
difficulty translates into a competition between the cylindrical
symmetry of the applied magnetic field, and the spherical symmetry
of the Coulomb interactions.
An often applied approximation is the adiabatic approximation\cite{schiff},
in which the electronic orbital is assumed to be a product of a
Landau level\cite{ruder} for the direction transverse to the
magnetic field (in $\rho^2=x^2+y^2$) and a longitudinal function (in $z$)
basically determined by the Coulomb interactions.

Several studies have recently addressed the electronic structure of helium
atoms subjected to strong magnetic fields.
Most works have examined only the lowest electronic states
using Hartree--Fock (HF)\cite{henry,vincke,ivanov} and 
variational\cite{larsen,scrinzi} methods.
To be predictive of observed stellar spectra, however, many excited
states are required, not only the lowest atomic state for a given
symmetry.
These excited states must also be determined with sufficiently high
accuracy to distinguish the dominant absorption features found in observed
stellar spectra.
Several recent attempts have been made to solve the HF self-
consistent field (SCF) equations for the spectrum of magnetized helium.
Thurner {\it et al}. \cite{thurner}
used numerical quadrature of the HF SCF equations to obtain results 
for several excited states of helium atoms and helium-like ions
over a range of magnetic fields up to $10^{12}\,\textrm{G}$.
The errors in this method, however, were best illustrated by later
calculations of Jones {\it et al}.\cite{jones96}
who applied a basis set of Slater-type orbitals (STO) in solving 
the same equations, and
were able to obtain lower energies for two excited states over many symmetries,
but only up to magnetic field strengths of $10^{10}\,\textrm{G}$.
Later quantum Monte Carlo (QMC) calculations\cite{rp},
using these same STO wave-functions as a starting point,
found that the residual basis set truncation errors were still significant
over much of the range of magnetic field strengths studied, and
emphasized the need for more accurate HF wave-functions.
The helium spectrum in strong magnetic fields is the primary focus of 
this work.
To more accurately determine the spectrum of neutral helium,
we elected to stay within the basis set HF approach and utilize
a much more flexible set of basis functions.

In this paper we use a basis set introduced by Aldrich and 
Greene\cite{aldrich} in combination with a systematic method for 
saturating the basis set\cite{rudenberg} to study the lowest energy
electronic states of neutral helium.
This basis set consists of functions of the form
\begin{equation}
\chi(\textbf{r})\propto \exp(-a\rho^2-bz^2),
\end{equation}
where $a$ and $b$ are variational parameters.
This basis set has the advantage that it can be used to accurately
represent states in which the charge density is highly anisotropic, 
where the values of the constants $a$ and $b$ differ.
Combined with our previous method\cite{jones96}
for obtaining excited state solutions to the HF equations, we obtain
three excitations for each of the symmetries studied, with a precision
that we estimate to be approximately $0.001$ atomic energy units (hartree).
We estimate the remaining basis set truncation error using our released--
phase QMC method\cite{rp} which, in principle, is able to obtain the 
exact energies, including electron correlation.
We also provide tables of dipole strengths calculated for the lowest
three excitations of each symmetry, a necessity for accurate modeling
of the absorptive behavior present in the atmospheres of strongly
magnetized stars.
In the first section we briefly review the methods employed in this
study.
The second section then presents our results, beginning with the
HF energies and dipole matrix elements, and we consider the 
implications of our results for the adiabatic approximation.
We also use stochastic methods to determine the correlation energy
and estimate the remaining basis set truncation error in the HF
energies.  
We conclude with some remarks about the applicability of our results 
to models of the atmospheres of magnetized compact stellar remnants.

\section{Method}
Our method, apart from the introduction of a different basis set,
is essentially the same as that of our previous paper\cite{jones96}.
Here we recapitulate only the essential formulae.
The magnetic field strength is parameterized by the constant
$\beta_{Z}=\beta/Z^{2}=B/B_{0}Z^2$, where 
$B_{0}=4.701\times 10^{9}\,\textrm{G}=4701\,\textrm{MG}$.
The Hamiltonian in atomic units for an atom with $N$ electrons and
atomic number $Z$ in constant magnetic field (along the $z$-direction)
is given by
\begin{eqnarray}
\widehat{\rm I\!H} &=& \sum_{i=1}^{N} \left [ -\frac{\nabla_i^2}{2}
- \frac{Z}{r_i} + \frac{(Z^2 \beta_Z)^2}{2} (x_i^2+y_i^2) \right]\cr
&& +Z^2\beta_Z \ ( L_z + 2 S_z) + \sum_{1 \leq i < j \leq N} \frac{1}{r_{ij}}
\;\;, \label{hamilt}
\end{eqnarray}
where $L_z = \sum_{i=1}^{N} \ell_{i z}$ and $S_z = \sum_{i=1}^{N}
s_{i z}$ are the $z$-component of the total angular momentum and spin
of the system, respectively,
and lengths are in units of the Bohr radius $a_0$.
Here we neglect contributions arising from the finite nuclear mass.
We have chosen the magnetic field to be parallel to
the $z$-axis, and the symmetric gauge, which has vector potential
${\bf A} = B(-y,x,0)/2$.
In the absence of external fields the eigenvalues of $L^2, L_z,
S^2, S_z$, and parity, $\Pi$, are good quantum numbers.
When the magnetic
field is turned on, the rotational invariance is broken and the only
conserved quantum numbers are the eigenvalues of $L_z, S^2, S_z$, and
$\Pi$ (alternatively, we will use the {\it z}-parity,
$\Pi_z$).
We will use both the zero field notation,
and also the triplet of proper quantum numbers $M$ (the 
eigenvalue of $L_z$), $\pi_z$ and $S_z$ in the form ($M,\pi_z,S_z$).
Taking our wave-function $\Psi$ to be a single Slater determinant and
minimizing the energy of the above Hamiltonian with respect to the electronic
spin orbitals, $\{ \psi_a \}$ [
$\psi_a({\bf x}) = \alpha(s) \otimes \phi_a({\bf r})$,
where $\alpha(s)$ is a spin function, $\phi_a({\bf r})$ a spatial
orbital, and ${\bf x}=(s,{\bf r})$],
we obtain the usual Hartree-Fock equations,
\begin{equation}
\label{eq:hf}
F \psi_a = \epsilon_a \psi_a,
\end{equation}
\noindent where $F$ is the single particle Fock operator,
\begin{equation}
F = h({\bf r}) +\sum_b \left( {\cal J}_b-{\cal K}_b \right),
\end{equation}
\noindent and,
\begin{eqnarray}
\nonumber h({\bf r}) &=& - \frac{1}{2} \nabla^2 -\frac{Z}{r}
 +\frac{(Z^2\beta_Z)^2}{2}\left( x^2 + y^2 \right) \cr
&& +Z^2\beta_Z
 \left( l_z + 2s_z \right), \\
\nonumber {\cal J}_b \psi_a &=& \left[ \int d{\bf x}' |{\bf r}-{\bf r}'|^{-1}
     \psi_b^{*} ({\bf x}')
     \psi_b({\bf x}') \right] \psi_a({\bf x}),\\
{\cal K}_b \psi_a &=& \left[ \int d{\bf x}' |{\bf r}-{\bf r}'|^{-1}
     \psi_b^{*}
({\bf x}') \psi_a({\bf x}') \right] \psi_b({\bf x}).
\end{eqnarray}
\noindent Note that we are still considering the integrals over the
spin degrees of freedom for the direct, ${\cal J}$, and exchange,
${\cal K}$, integrals.
We expand each spatial electronic orbital in a basis set,
$\{ \chi_\mu({\bf r}) \}$, of our choosing,
\begin{equation}
\phi_a({\bf r}) = \sum_{\mu=1}^{N_b} c_{a \mu} \chi_\mu({\bf r}),
\label{eq:orb}
\end{equation}
\noindent where $N_b$ is the number of basis set elements.
We have chosen to use the basis set of Aldrich and Greene\cite{aldrich}
\begin{equation}
\chi_\mu(\rho,\varphi,z) = N_\mu \ \rho^{|m_\mu|} \ z^{p_\mu} \ 
   e^{-im_\mu\varphi} e^{-a_\mu\rho^2-b_\mu z^2},
\label{eq:adg}
\end{equation}
where $m_\mu$ denotes the angular momentum quantum number of operator
$l_z$, and $p_\mu=0$ for positive
$z$-parity states, and $p_\mu=1$ for negative $z$-parity states.
The parameters $a_\mu$ and $b_\mu$ allow for different treatment of the 
transverse and longitudinal distance dependence, a crucial 
consideration when the applied magnetic field gets strong enough that 
the atom tries to minimize the diamagnetic contribution to the total
energy.  
Analytic expressions can be worked out for most of the matrix elements,
while the nuclear repulsion\cite{aldrich} and electron-electron matrix 
elements can be reduced to one dimensional integrals that are performed 
numerically.
Our expressions for the electron--electron matrix elements can be found
in the Appendix.

Our present calculations restrict the orbitals to have a common spatial
dependence for two electrons in the same state but of opposite spin.
In other words we use the restricted Hartree--Fock (RHF) approach.
Details of how we achieve the solution in this method can be found
in Ref. \cite{jones96}.

\subsection{Even-Tempered Gaussians}
An important concern in Eq. \ref{eq:orb} is whether or not the basis
set is sufficiently well converged for a desired level of accuracy
in the atomic total energy.
Previous calculations\cite{jones96} have been plagued by an
inadequate basis set - even the largest basis sets used have not been 
well converged in a systematic way.
In this work we have used even-tempered Gaussian (ETG) 
sequences\cite{rudenberg}
to systematically saturate the basis set for each electronic orbital.
Essentially this means that a sequence of $N_s$ $a$ and $b$ parameters 
are generated
in such a way that they fill the possible range of values as each sequence
is made longer ($N_s$ increases).
In zero field calculations only one sequence of spheroidal Gaussians
is needed for each angular momentum type (for Gaussian type orbitals,
$a=b$, and the Gaussian is multiplied by a spherical harmonic).
These sequences of basis elements are given by
\begin{equation}
b_i = b_1 \cdot \delta_b^{i-1}, \qquad i=1,\ldots,N_s,
\label{eq:etg_b}
\end{equation}
where $b_1$ and $\delta_b$ are the variational parameters for the
entire sequence that must be optimized with respect to the HF total energy.
As $N_s$ gets longer, the basis set saturates, or becomes more and more
complete.
These sequences have been successfully used for well saturated
zero field atomic calculations\cite{rudenberg}, a case in which the
parameters $a$ and $b$ are equal.
It should be emphasized that these ETG sequences are simply a reliable
way in which the basis set can be saturated.
In practice, we fully optimize the values of the basis set parameters
for each value of $N_s$, the length of the sequences, with respect to 
the total energy.  One can then also fit the resulting optimal 
parameters in order to obtain a scheme for extrapolating to even 
larger basis set sizes (sequence lengths)\cite{rudenberg}.
However, we have not done this extrapolation procedure, as the
optimal sequences are sufficiently well converged in energy to meet 
our requirements for the accuracy of the total energy.

The main complication in the present application is that our basis
set (Eq. \ref{eq:adg}) is designed to break spherical symmetry ($a\ne b$),
such that we have two parameters in each Gaussian basis function,
separately describing the longitudinal and transverse directions.
To compensate for this separation, we use a series of such 
ETG sequences, with the longitudinal parameters given by
Eq. \ref{eq:etg_b}, and the matching transverse parameters given by
\begin{equation}
a_i = f\cdot b_i, \qquad i=1,\ldots,N_{s},
\end{equation}
where $f$ is a constant factor.
We have typically selected $f$ such that we use a series of $2-5$ even-tempered
sequences for each orbital, with $f=1,2,4,8,\ldots$.
Note that the total number of basis functions is $N_s$ multiplied by 
the number of different sequences.
Additionally we have included a sequence such that $a$ is fixed, $a=a_B$,
where $a_B=Z^2\beta_{Z}/2$ is the exponential parameter for the first Landau 
level of an electron in a constant magnetic field.
As the magnetic field gets larger, this Landau level sequence
becomes more and more important in the basis set expansion of the
electronic orbitals.
The adiabatic approximation corresponds to using only sequences in 
which $a$ is fixed to $a_{B}$.
We will explore this issue further below.

\section{Results}

\begin{figure}[t]
\centering\epsfig{file=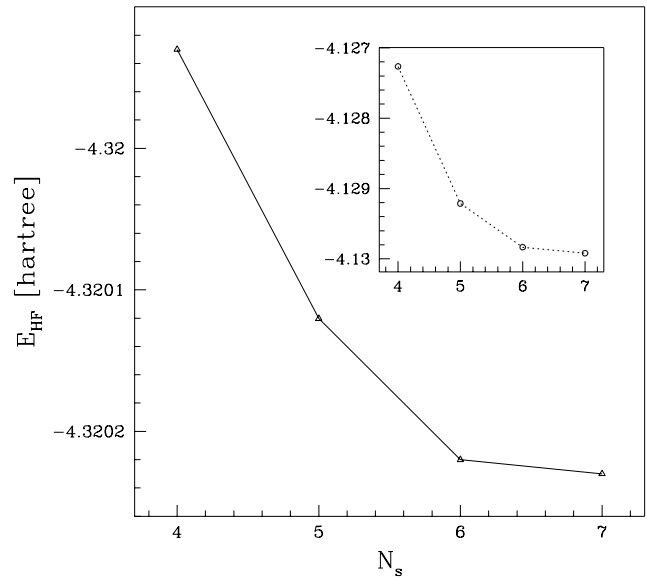,height=3.5in}
\caption{Sample convergence of the first and third (inset) states of
$(M,\pi_z,S_z)=(0,+,-1)$ symmetry (1s2s and 1s4s at zero field) at 
$\beta_Z=1$.}\label{fig:converge}
\end{figure}

Our desired accuracy for the HF total energies is $0.001$ hartree,
such that we can resolve transitions in the optical regime with
a precision of $\delta\lambda \lesssim 100$\AA, where
$\lambda= hc/|E_i-E_f|$ is the wavelength of allowed transitions between
initial and final states.
Our RHF calculations reported here used a minimum of $2$ ETG
sequences for each orbital, with some calculations
using up to $5$ sequences.
The convergence of the total energy as a function of the number of 
sequences was carefully checked at several different magnetic field
strengths (typically $\beta_Z=1,10,100$), and was converged
to within $0.0005$ hartree.
It was necessary to include the Landau level sequence ($a_i = a_B$)
for the 1s orbital for $\beta_Z\gtrsim 1$, while it was always
necessary (apart from $\beta_Z=0$) to include the Landau level 
sequence for the second electronic orbital, due to the much
greater impact of the applied field on the outermost electron.
We began calculations with each sequence having a length of four
($N_s=4$ in Eq. \ref{eq:etg_b}) and fully optimized the parameters
for each sequence (two parameters for each sequence).
This process was repeated until the energy converged within
$0.0001$ hartree, which typically required a length of $N_s=6-8$.
A typical example of this behavior is shown in Figure \ref{fig:converge}
for the first and third (inset) states of symmetry $(M,\pi_z,S_z)=(0,+,-1)$.
The total energy in this example is amply converged by $N_s=7$.
Note that we have chosen more stringent cut-offs for the convergence
of the total energy with respect to the number and length of ETG
sequences; thus the desired accuracy of 0.001 hartree is a conservative
estimate of the remaining basis set truncation error.

\subsection{Excited State Spectrum of Neutral He}
Tables \ref{tab:he1}-\ref{tab:he8} contain the HF-ETG energies computed
according to the method outlined above, for the spin-polarized ($S_z=-1$)
symmetries having $M=0,-1,-2,-3$ and $\pi_z=\pm$.
We note that these energies are always lower than the best previously
published HF results\cite{jones96}, with the exception of some of the 
very lowest states of each symmetry at small applied magnetic field
strength.
This slight degradation (generally around $0-3\times 10^{-4}$ hartree)
is due to the ETG basis functions not representing the correct
cusp behavior\cite{kato} at the nucleus, while the Slater-type
orbitals do possess a nonzero derivative at the origin.
The ETG basis elements
are much better, however, at reproducing the highly anisotropic behavior
at high magnetic fields.
This improved accuracy is reflected in the fact that the higher 
excited states are much superior to those published previously\cite{jones96}.

\begin{figure}[t]
\centering\epsfig{file=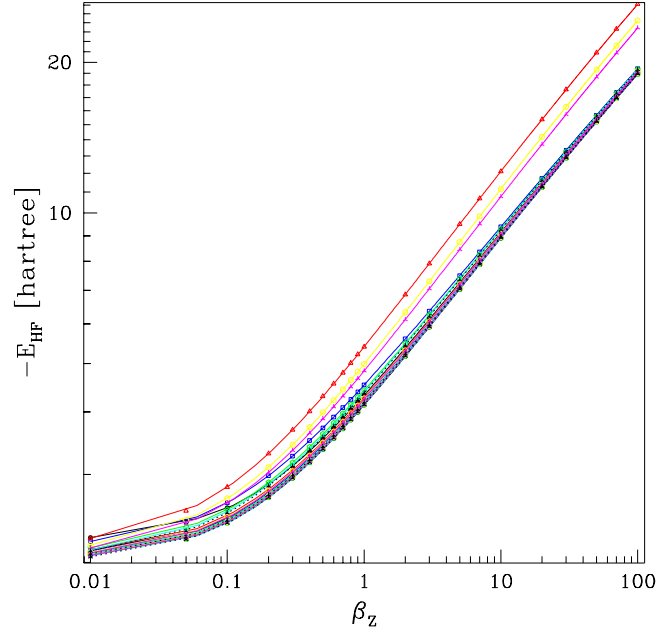,height=3.5in}
\caption{The energy spectrum of 24 states of neutral helium in applied
longitudinal magnetic fields up to $\beta_Z=100$, or approximately
$2\times{10^{12}}\,{\rm G}$.}\label{fig:all_en}
\end{figure}

The spectrum of the energy states computed thus far is shown in Figure
\ref{fig:all_en}.
We note that the most tightly bound states (the first states of 
$(-1,+,-1)$,$(-2,+,-1)$,$(-3,+,-1)$ symmetry) lie
lowest in energy, by a considerable margin, with a gap to higher
excitations that increases monotonically with applied field strength.
The remaining states form a broad band of excitations whose width is
slowly increasing as a function of magnetic field strength.
It should be noted that many additional tightly bound states
lie between the band of excitations and the lowest tightly bound 
$(-3,+,-1)$ state, for example, $(-4,+,-1)$,$(-5,+,-1)$,$\ldots$.
We have chosen to focus on the most energetically favorable states
at high magnetic fields.
Thus we have considered only spin triplet states,
and have restricted ourselves to states with $-3\le M\le 0$.
If it is necessary to compute more symmetries to accurately model
magnetized white dwarf atmospheres, it is a simple matter to extend
the calculations presented in this work.

\subsection{Dipole Strengths for Low Lying He Transitions}
\begin{figure}[t]
\centering\epsfig{file=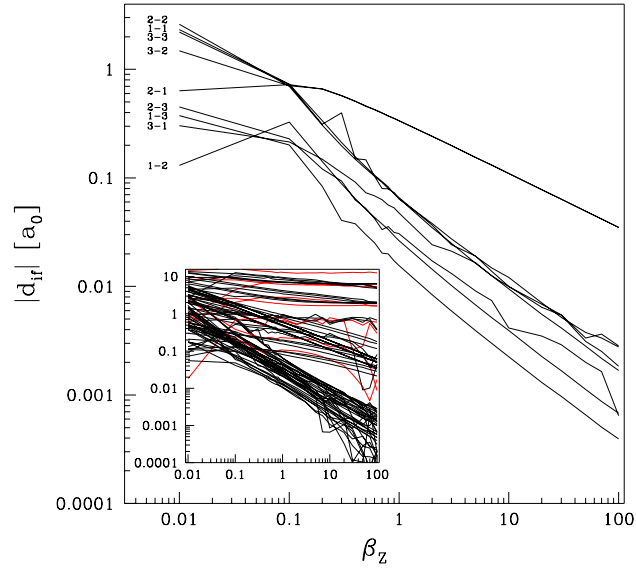,height=3.5in}
\caption{Dipole strengths for allowed transitions in neutral helium as
a function of magnetic field strength.
The main figure shows the transitions from symmetry $(M,\pi_z,S_z)=(-1,+,-1)$
to $(0,+,-1)$, with labels denoting the level of
excitation.
For example, 1-1 is the transition between zero field
states ${\rm 1s2p}_{-1}$ and ${\rm 1s2s}$.
The inset shows all of the data for all eight symmetries
that we have considered, for a total of ninety transitions.}\label{fig:all_dip}
\end{figure}

The dipole matrix element, in atomic units, between initial state $\Psi_i$
and final state $\Psi_f$ is given by
\begin{equation}
d_{if} = \sum_{j=1}^2 \left\langle \Psi_i \left| \sqrt{\frac{4\pi}{3}} r_j 
    Y_{1,\Delta M} \right| \Psi_f \right\rangle,
\end{equation}
where $\Delta M = M_f - M_i$, and $Y_{1,\Delta M}$ is the usual
spherical harmonic.
These dipole matrix elements vanish unless the zero field angular
momentum quantum numbers (eigenvalues of the operator $L^2$)
differ by one, $|L_f-L_i|=1$.
This same rule also applies at nonzero applied field, as the
diamagnetic term only couples states that differ by two
in $L$.
These selection rules allow for ten possible transitions
between states of different symmetry (that we have considered).
Numerical tables for the dipole matrix elements can be provided by 
the authors upon request, or obtained on the World Wide Web
({\tt http://www.ncsa.uiuc.edu/Apps/CMP/papers/jon98/jon98.html}).
Graphical results for the dipole matrix elements are shown in 
Figure \ref{fig:all_dip}.
The ninety allowed transitions plotted in Figure \ref{fig:all_dip} show some
basis set truncation error, which is not unexpected, as the HF-SCF 
wave-functions
are optimized according to energy, leaving other expectation values
more sensitive to the basis set error.

\subsection{Validity of the Adiabatic Approximation}
\begin{figure}[t]
\centering\epsfig{file=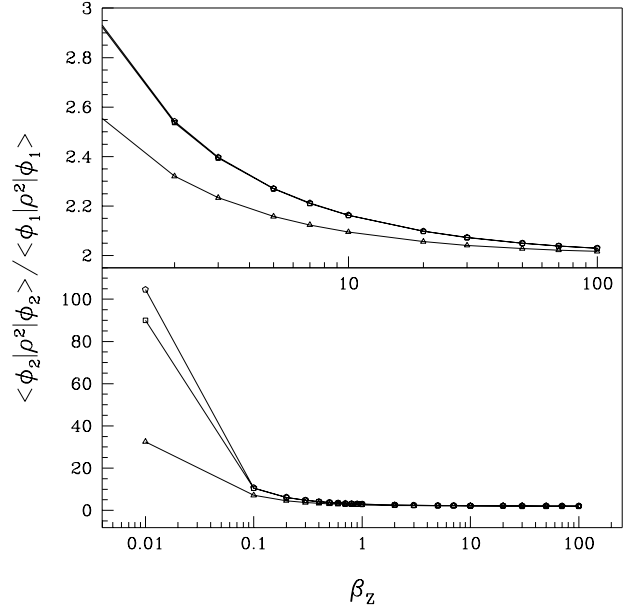,height=3.5in}
\caption{The ratio of the expectation value of $\rho^2$ for the second 
electronic orbital relative to the first for the first three excited states 
of $(M,\pi_z,S_z)=(-1,+,-1)$ symmetry.
The upper panel is a close-up of the data for larger field strengths.
The ratio should approach 2.0 in the adiabatic limit.
Lines are provided only as a guide to the eye.}\label{fig:p2}
\end{figure}

In the often used adiabatic approximation\cite{schiff}, the only basis 
functions used are those corresponding to the lowest Landau level, while
the functional dependence in the longitudinal direction is allowed
to vary.
We assess the validity of such an approximation by measuring the
relative average transverse ``widths'' of the electron orbitals, namely
\begin{equation}
\left\langle \phi_2 \left| x^2+y^2 \right| \phi_2 \right\rangle /
	\left\langle \phi_1 \left| x^2+y^2 \right| \phi_1 \right\rangle
=\frac{\left\langle\rho^2\right\rangle_2}{\left\langle\rho^2\right\rangle_1},
\end{equation}
where $\phi_1$ corresponds to the first (1s) orbital, and $\phi_2$ the
second, excited state, orbital.
If the adiabatic approximation is valid, this ratio of the widths
of the second electron compared to the first should approach a
constant (see Eq. \ref{eq:adlim}).
We show examples of this width for the ${\rm 1s2p}_{-1}$,
${\rm 1s3p}_{-1}$ and ${\rm 1s4p}_{-1}$ states in Fig. \ref{fig:p2}.
The lower panel shows the width for all values of the applied 
magnetic field, while the upper panel focuses on the region where
$\beta_Z \gtrsim 1$.
We note that the ratio only approaches a constant for the very 
largest field strengths, $\beta_Z \gtrsim 80-100$, and
is still slowly varying even in this super-strong field regime.
The ratio is also only slightly dependent on the degree of excitation.

\begin{figure}[t]
\centering\epsfig{file=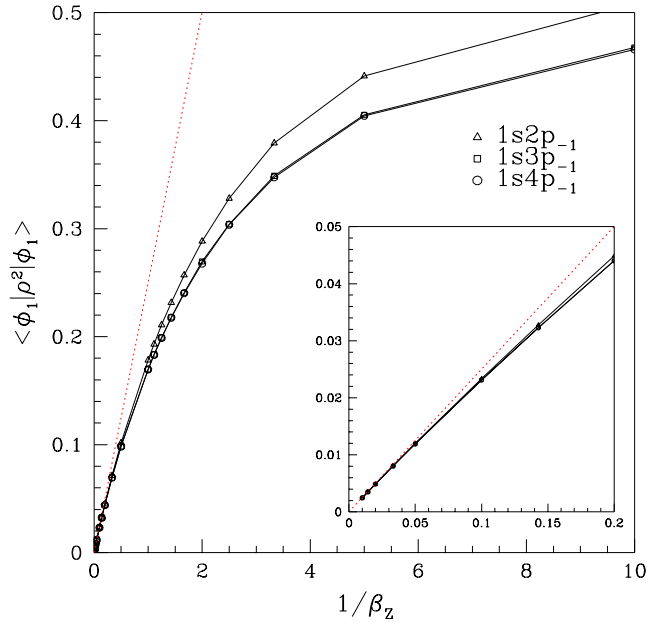,height=3.5in}
\caption{The expectation value of $\rho^2$ for the first HF electronic
orbital $\phi_1$ as a function of magnetic field strength, which
should become linear in the limit when the adiabatic approximation is
valid.  Three orbitals are shown, corresponding to the first three
excitations of $(M,\pi_z,S_z)=(-1,+,-1)$ symmetry.
The inset shows a close-up of the super-strong field regime
near the origin.  The dashed line indicates the adiabatic limit.
Solid lines are provided only as a guide to the eye.}\label{fig:rho1}
\end{figure}

\begin{figure}[t]
\centering\epsfig{file=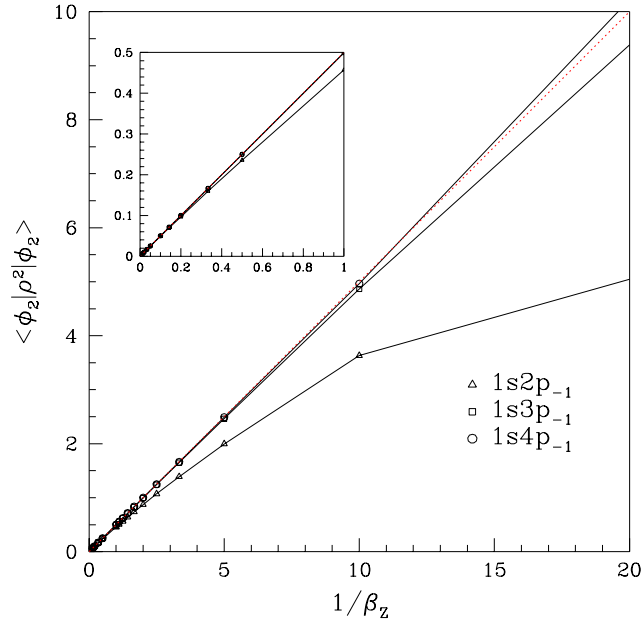,height=3.5in}
\caption{The expectation value of $\rho^2$ for the second (excited) HF 
electronic orbital $\phi_2$ as a function of magnetic field strength, which
should become linear in the limit when the adiabatic approximation is
valid.  Three orbitals are shown, corresponding to the first three
excitations of $(M,\pi_z,S_z)=(-1,+,-1)$ symmetry.
The inset shows a close-up of the super-strong field regime
near the origin.  Solid lines are provided only as a guide to the eye.
The dashed line indicates the adiabatic limit.}\label{fig:rho2}
\end{figure}

We can learn more by examining separately the behavior of
each of the two electronic orbitals.
In the limit of infinite magnetic field strength, the electrons
should occupy the lowest Landau level,
\begin{equation}
\Phi_{nm}^{Lan}(\rho,\varphi)=\frac{\sqrt{n!}}{\sqrt{2\pi(n+|m|)!l^2}}
 \left(\frac{\rho}{\sqrt{2}l}\right)^{|m|} {\rm L}_n^{|m|}
 \left(\frac{\rho^2}{l^2}\right)e^{-im\varphi}e^{-\rho^2/4l^2},
\end{equation}
where $l=\sqrt{1/2\beta}$ is the magnetic length, and ${\rm L}_n^{|m|}$ is
an associated Laguerre polynomial.
For the lowest Landau state, $n=0$, and we have
\begin{equation}
\Phi_{0m}^{Lan}=\left[\frac{2}{2\pi|m|!}\right]^{\frac{1}{2}}
\beta^{\frac{|m|+1}{2}}\rho^{|m|}e^{-\beta\rho^2/2}e^{-im\varphi}.
\end{equation}
The full wave-function also includes a factor of an unknown function
of $z$ times the cylindrical Landau state, but we are concerned
here with the quality of the adiabatic approximation, so we focus
only on the accuracy of the description of the transverse behavior.
Now we consider the expectation value of $\rho^2$ in the lowest Landau
orbitals,
\begin{equation}
\label{eq:adlim}
\left\langle\Phi^{Lan}_{0m}\left|\rho^2\right|\Phi^{Lan}_{0m}\right\rangle
  = \frac{|m|+1}{\beta}=\frac{|m|+1}{Z^2\beta_Z}.
\end{equation}
We see that, in the adiabatic limit, the expectation value of $\rho^2$
for each electronic orbital should be a simple constant divided by
the magnetic field strength.
Figures \ref{fig:rho1} and \ref{fig:rho2} plot this expectation value
as a function of $1/\beta_Z$ for the same example states that we considered
above, the ${\rm 1s2p}_{-1}$, ${\rm 1s3p}_{-1}$ and ${\rm 1s4p}_{-1}$ states.
Figure \ref{fig:rho1} plots $\langle\rho^2\rangle$ for the first 
electronic orbital
of helium (the 1s state at zero magnetic field), while Figure \ref{fig:rho2}
shows the same result for the second orbital.
We see in Figure \ref{fig:rho1} that the approach to the adiabatic limit
(indicated by the dashed line, whose slope is $1/4$ by Eq. \ref{eq:adlim})
is really not valid until $\beta_Z\gtrsim 50$, a very large field
indeed (about $10^{12}\,\textrm{G}$).
We also note that the limit is reached at approximately the same
value of field strength for all three states, a reasonable result,
as the innermost electron should not be greatly different from
one excited state to the next.
The second orbital, however, reaches the adiabatic limit much more quickly,
as we see from Figure \ref{fig:rho2}.
The more spatially extended states feel a much larger effective
magnetic field (due to the diamagnetic term in the Hamiltonian);
thus the tightly bound ${\rm 2p}_{-1}$ state (given by the triangles)
reaches the adiabatic limit (indicated by the dashed line of slope $1/2$)
more slowly ($\beta_Z\simeq 5$) than the next two excited states, 
which are very near the adiabatic regime at $\beta_Z\simeq 0.1-0.2$.
The implications of these results for multi-electron atoms is that
the adiabatic approximation is not good for the innermost electrons,
except at extremely large field strengths, due to the importance of
the Coulomb repulsion from the nucleus.
Thus we conclude that the adiabatic approximation for multi-electron atoms
is seldom very accurate, even for the large magnetic fields found
in magnetized white dwarfs and neutron stars, which is generally
less than $10^{12}\,\textrm{G}$.

\subsection{Basis Set Truncation Error and Correlation Energy}
\begin{figure}[t]
\centering\epsfig{file=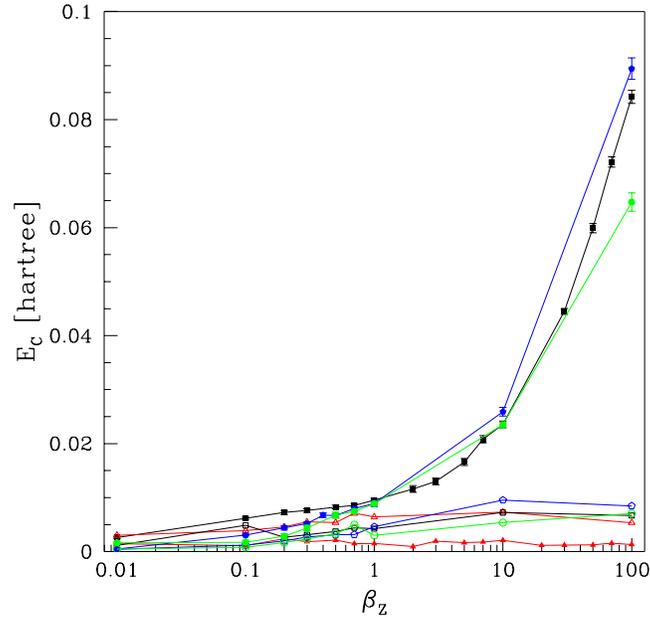,height=3.5in}
\caption{The correlation energy from FPQMC as a function of magnetic field 
strength for the
lowest state of each symmetry corresponding to $S_z=-1$, $M=0$ (triangles),
$M=-1$ (squares), $M=-2$ (pentagons), and $M=-3$ (circles).
The positive $z$-parity states have solid symbols.
Lines are provided only as guides to the eye.}\label{fig:Ec}
\end{figure}

To provide an estimation of the size of our basis set truncation error,
we have used two quantum Monte Carlo methods.
The fixed--phase method\cite{ortiz} (FPQMC) is a variational method that 
projects out the ground state of a particular symmetry using
stochastic random walks.
If the state were bosonic, FPQMC would yield an exact result (albeit
with statistical error bars) for the total energy.
Since we have fermions, the FPQMC energies are an upper bound to the
exact total energy (often a very good upper bound) for the ground state
of a given symmetry, whose quality
is constrained by the fixed--phase approximation.
The released--phase method\cite{rp} (RPQMC) is an ``exact'' method that 
can simultaneously determine the ground and excited states by using 
correlation functions in imaginary time.
If sufficiently well converged in imaginary time, this RPQMC method
obtains the exact energies, but always provides at least an upper
bound to the exact excited state energies.
Both of these methods, along with an earlier application to low-lying excited
states of helium (for smaller and somewhat inferior STO basis sets),
are reviewed in Ref. \cite{rev}.
It is difficult, of course, to separate out the basis set truncation errors
from the correlation energy,
\begin{equation}
E_C = E_{HF} - E_{QMC}.
\end{equation}
Our ETG basis set should be equally valid at all field strengths,
hence we can look for basis set error by examining the behavior of the
correlation energy as a function of field strength.
Errors arising from truncation of the basis set should show up
as (small) perturbations on the otherwise smooth correlation energy
curve.
The behavior of the correlation energy for the lowest state of each 
of the eight symmetries studied is shown in Fig. \ref{fig:Ec}.
From the small oscillations in the correlation energy curves
we estimate, for the range of magnetic field 
strengths $0\le\beta_Z\le 1$, that the truncation error is generally less
than $0.001$ hartree.
For larger fields the truncation error increases.
We note that $E_C$ increases slowly with applied field strength,
except for the most tightly bound states, which have zero field quantum 
numbers ${\rm 1s2p}_{-1}$, ${\rm 1s3d}_{-2}$,
and ${\rm 1s4f}_{-3}$, which increase rather dramatically as the field
strength grows larger.

Table \ref{tab:rp} compares our current RPQMC results for a selected
set of magnetic field strengths and all eight symmetries studied with
our previous RPQMC results\cite{rp}.
We note that the previous results suffered from poor wave-function quality,
as the present results (the method, when insufficiently converged in
imaginary time, remains variational) are greatly improved, especially
for the higher excitations.
We also note that the correlation energy is quite small for the
highest excitations, regardless of the symmetry state.
This reduction in correlation energy for the highest excitations
is most likely due to the large physical separation between the
innermost and outermost electrons.

\subsection{Comparison with Other Calculations}

For our HF-ETG results we have already noted a favorable comparison
with the best HF excited state calculations in the literature\cite{jones96}.
For our fully correlated RPQMC results we compare, in Table \ref{tab:rp_comp},
with the recent work of Scrinzi\cite{scrinzi}, who applied a
variational calculation with a correlated basis to the first three
excitations of $M=0$ and $M=-1$, and Becken {\it et al}.\cite{becken},
who performed a very large configuration interaction (CI) calculation
for the $M=0$ symmetries, with up to six excited states.
We note that our results compare favorably with the calculations
of Becken {\it et al}., at least for the symmetries that they have
computed thus far, while there are large discrepancies with the
results of Scrinzi.
Some of Scrinzi's values for the total energy are considerably lower in
energy than both our RPQMC results and the CI values.
The agreement between the distinctly different CI and RPQMC methods is
reassuring, and
it seems most likely that these anomalously low energies 
of Scrinzi reflect numerical errors.

\section{Conclusions}
We have applied a systematic method of basis-set saturation within the
Hartree--Fock formalism for the case of neutral helium in strong magnetic 
fields,
obtaining an accuracy of approximately $0.001$ hartree atomic units
over a wide range of magnetic field strengths.
The resulting accuracy in determining wavelengths for transitions
among these states is thus 
$\delta\lambda/\lambda \lesssim 0.0021/\Delta E_{HF}$, for 
$\beta_Z \lesssim 1$.
For optical transitions, this accuracy is  $\lesssim 2\%$.
Unfortunately, the dipole matrix elements also have truncation errors,
which are much more difficult to estimate.

Using our accurate HF-ETG wave-functions, we have used quantum Monte
Carlo methods to determine the correlation energy (the difference between
Hartree--Fock and the exact total energy) and estimate the
residual basis set truncation energies.
We have also evaluated the validity of the adiabatic approximation,
and found that it is poor for the lowest lying states and
magnetic field strengths.
For example, the ground state of neutral helium does not enter the 
adiabatic regime until $\beta_Z \gtrsim 50$.

Both the dipole strengths and transition energies are required to
construct a detailed model of the atmospheres of magnetic white dwarfs,
which has not yet been done for stars suspected of containing neutral helium.
We hope that the extensive tabulations provided in this work can
provide meaningful input into such models.
Tables of numerical results for both the dipole matrix elements and
energies can be obtained on the World Wide Web at 
URL:{\tt http://www.ncsa.uiuc.edu/Apps/CMP/papers/jon98/jon98.html}.

\acknowledgements
Calculations were performed at the National Center for Supercomputing
Applications and the Cornell Theory Center.
This work was performed under the auspices of the U.S. Department of
Energy.

\appendix
\section*{Electron--Electron Matrix Element}
\label{app1}

The strategy that we have used for the electron--electron matrix elements
reduces the six dimensional integral to a one dimensional one that can 
be rapidly evaluated using standard numerical quadrature methods.
The matrix elements between basis functions of the form Eq. \ref{eq:adg},
\begin{equation}
I_{\mu\nu\lambda\sigma}^{ee} = \left\langle \chi_\mu({\bf r}_1)
  \chi_\lambda({\bf r}_2) \left| \frac{1}{\left|{\bf r}_1-{\bf r}_2\right|}
  \right|\chi_\nu({\bf r}_1)\chi_\sigma({\bf r}_2)\right\rangle,
\end{equation}
can be expanded using the identity
\begin{equation}
\frac{1}{r_{12}}=\frac{1}{\left|{\bf r}_1-{\bf r}_2\right|}=
\frac{1}{2\pi^2}\int_0^\infty du\int_{\rm I\!R^3}{d{\bf k}}
  e^{i{\bf k}\cdot{\bf r}_{12}-k^2u}.
\end{equation}
We now expand the three dimensional integral over ${\bf k}$ in 
cylindrical coordinates $k_\rho$,
$k_z$ and $k_\varphi$,
\begin{equation}
I_{\mu\nu\lambda\sigma}^{ee} = \delta_{m_1 m_2} N_{\mu\nu\lambda\sigma}
\int_0^\infty du I_{k_\rho}(u) I_{k_z}(u),
\end{equation}
where $N_{\mu\nu\lambda\sigma}=N_\mu N_\nu N_\lambda N_\sigma$,
$m_1=m_\mu+m_\nu$, $m_2=m_\lambda+m_\sigma$, and the remaining
one dimensional integral is evaluated numerically.
The two expressions remaining in the integrand are the results from the
integration over $k_\rho$ and $k_z$.
\begin{align}
I_{k_\rho}(u) =& \frac{\pi}{2^{2m_1+1}}
             \frac{(n_1+n_2+m_1)!}{a_1^{n_1+m_1+1}a_2^{n_2+m_1+1}}
  \frac{(u+\frac{1}{4a_1})^{n_2}(u+\frac{1}{4a_2})^{n_1}}
     {(u+A)^{n_1+n_2+m_1+1}} \nonumber\\
 &\quad \times F\left(-n_2,-n_1;-n_2-n_1-m_1;
\frac{u(u+A)}{(u+\frac{1}{4a_1})(u+\frac{1}{4a_2})}\right),
\end{align}
where $A=1/[4(a_1+a_2)]$ and $F$ is the confluent hyper-geometric function, 
in this particular case a relatively simple finite series.
\begin{align}
I_{k_z}(u) &= (-1)^p\frac{\pi^{3/2}}{8^p}b_1^{-p_1-1/2}b_2^{-p_2-1/2}p_1!p_2!
  (u+C)^{-p-1/2} \nonumber\\
      & \quad\times \sum_{j_1=0}^{\left[\frac{p_1}{2}\right]}
      \sum_{j_2=0}^{\left[\frac{p_2}{2}\right]}
  \frac{(-1)^{j_1+j_2}b_1^{j_1}b_2^{j_2}(2p-2j_1-2j_2-1)!!}{j_1!j_2!(p_1-2j_1)!
    (p_2-2j_2)!}
  \left[2\left(u+C\right)\right]^{j_1+j_2},
\end{align}
where $C=1/[4(b_1+b_2)]$ and the numbered quantities are related to the 
parameters of the original basis functions by
\begin{gather}
a_1 = a_\mu + a_\nu \qquad a_2 = a_\lambda + a_\sigma,\\
b_1 = b_\mu+b_\nu \qquad b_2 = b_\lambda+b_\sigma,\\
p_1 = p_\mu + p_\nu \qquad p_2 = p_\lambda + p_\sigma, \\
n_1 = (|m_\mu|+|m_\nu|-m_\mu-m_\nu)/2 \qquad n_2 
  =(|m_\lambda|+|m_\sigma|-m_\lambda-m_\sigma)/2, 
\end{gather}
and $p_1+p_2=2p$ must be even (otherwise the integral 
$I_{\mu\nu\lambda\sigma}^{ee}$ is zero).

%

\vfill\eject
%
\begin{table}[t]
\caption{HF-ETG total energies for the first three excited states of
$(M,\pi_z,S_z)=(0,+,-1)$ symmetry.
Zero field quantum numbers are given at the top of each column.
}\label{tab:he1}
\begin{tabular}{rrrr}
$\beta_Z$&${\rm 1s2s}$&${\rm 1s3s}$&${\rm 1s4s}$\\
\hline
0.01 & -2.2425 & -2.1199 & -2.0825 \\ 
0.05 & -2.4111 & -2.2789 & -2.2354 \\ 
0.10 & -2.5720 & -2.4425 & -2.4034 \\ 
0.20 & -2.8659 & -2.7363 & -2.6986 \\ 
0.30 & -3.1211 & -2.9879 & -2.9498 \\ 
0.40 & -3.3444 & -3.2078 & -3.1696 \\ 
0.50 & -3.5435 & -3.4041 & -3.3653 \\ 
0.60 & -3.7239 & -3.5820 & -3.5428 \\ 
0.70 & -3.8894 & -3.7453 & -3.7054 \\ 
0.80 & -4.0427 & -3.8967 & -3.8567 \\ 
0.90 & -4.1857 & -4.0379 & -3.9978 \\ 
1.00 & -4.3202 & -4.1708 & -4.1299 \\ 
2.00 & -5.3680 & -5.2080 & -5.1656 \\ 
3.00 & -6.1260 & -5.9592 & -5.9161 \\ 
5.00 & -7.2543 & -7.0792 & -7.0349 \\ 
7.00 & -8.1147 & -7.9344 & -7.8891 \\ 
10.00 & -9.1385 & -8.9525 & -8.9061 \\ 
20.00 & -11.4945 & -11.2975 & -11.2494 \\ 
30.00 & -13.1223 & -12.9189 & -12.8703 \\ 
50.00 & -15.4668 & -15.2553 & -15.2057 \\ 
70.00 & -17.2046 & -16.9879 & -16.9382 \\ 
100.00 & -19.2286 & -19.0072 & -18.9543 
\end{tabular}
\end{table}
\begin{table}[t]
\caption{HF-ETG total energies for the first three excited states of
$(0,-,-1)$ symmetry.
Zero field quantum numbers are given at the top of each column.
}\label{tab:he2}
\begin{tabular}{rrrr}
$\beta_Z$&${\rm 1s2p}_0$&${\rm 1s3p}_0$&${\rm 1s4p}_0$\\
\hline
0.01 & -2.2031 & -2.1105 & -2.0724 \\ 
0.05 & -2.4197 & -2.2747 & -2.2330 \\ 
0.10 & -2.6347 & -2.4537 & -2.4067 \\ 
0.20 & -2.9827 & -2.7592 & -2.7060 \\ 
0.30 & -3.2657 & -3.0155 & -2.9587 \\ 
0.40 & -3.5068 & -3.2377 & -3.1784 \\ 
0.50 & -3.7186 & -3.4355 & -3.3740 \\ 
0.60 & -3.9082 & -3.6144 & -3.5512 \\ 
0.70 & -4.0808 & -3.7786 & -3.7137 \\ 
0.80 & -4.2394 & -3.9303 & -3.8644 \\ 
0.90 & -4.3864 & -4.0725 & -4.0052 \\ 
1.00 & -4.5282 & -4.2056 & -4.1417 \\ 
2.00 & -5.5978 & -5.2446 & -5.1774 \\ 
3.00 & -6.3621 & -5.9965 & -5.9278 \\ 
5.00 & -7.4932 & -7.1163 & -7.0465 \\ 
7.00 & -8.3523 & -7.9708 & -7.9004 \\ 
10.00 & -9.3724 & -8.9875 & -8.9169 \\ 
20.00 & -11.7170 & -11.3306 & -11.2599 \\ 
30.00 & -13.3364 & -12.9503 & -12.8802 \\ 
50.00 & -15.6691 & -15.2855 & -15.2150 \\ 
70.00 & -17.3996 & -17.0169 & -16.9469 \\ 
100.00 & -19.4150 & -19.0335 & -18.9634 
\end{tabular}
\end{table}
\begin{table}[t]
\caption{HF-ETG total energies for the first three excited states of
$(-1,+,-1)$ symmetry.
Zero field quantum numbers are given at the top of each column.
}\label{tab:he3}
\begin{tabular}{rrrr}
$\beta_Z$&${\rm 1s2p}_{-1}$&${\rm 1s3p}_{-1}$&${\rm 1s4p}_{-1}$\\
\hline
0.01 & -2.2353 & -2.1220 & -2.0888 \\ 
0.05 & -2.5366 & -2.2972 & -2.2413 \\ 
0.10 & -2.8301 & -2.4855 & -2.4178 \\ 
0.20 & -3.3016 & -2.8006 & -2.7188 \\ 
0.30 & -3.6842 & -3.0630 & -2.9735 \\ 
0.40 & -4.0107 & -3.2897 & -3.1938 \\ 
0.50 & -4.2980 & -3.4903 & -3.3909 \\ 
0.60 & -4.5560 & -3.6743 & -3.5707 \\ 
0.70 & -4.7911 & -3.8411 & -3.7343 \\ 
0.80 & -5.0078 & -3.9954 & -3.8858 \\ 
0.90 & -5.2107 & -4.1388 & -4.0276 \\ 
1.00 & -5.4000 & -4.2742 & -4.1605 \\ 
2.00 & -6.8666 & -5.3261 & -5.1995 \\ 
3.00 & -7.9213 & -6.0856 & -5.9508 \\ 
5.00 & -9.4882 & -7.2156 & -7.0714 \\ 
7.00 & -10.6816 & -8.0768 & -7.9268 \\ 
10.00 & -12.1011 & -9.1010 & -8.9451 \\ 
20.00 & -15.3690 & -11.4589 & -11.2906 \\ 
30.00 & -17.6289 & -13.0873 & -12.9120 \\ 
50.00 & -20.8876 & -15.4327 & -15.2490 \\ 
70.00 & -23.3066 & -17.1710 & -16.9822 \\ 
100.00 & -26.1264 & -19.1943 & -18.9999 
\end{tabular}
\end{table}
\begin{table}[t]
\caption{HF-ETG total energies for the first three excited states of
$(-1,-,-1)$ symmetry.
Zero field quantum numbers are given at the top of each column.
}\label{tab:he4}
\begin{tabular}{rrrr}
$\beta_Z$&${\rm 1s3d}_{-1}$&${\rm 1s4d}_{-1}$&${\rm 1s5d}_{-1}$\\
\hline
0.01 & -2.1403 & -2.0903 & -2.0673 \\ 
0.05 & -2.3512 & -2.2588 & -2.2271 \\ 
0.10 & -2.5528 & -2.4382 & -2.4012 \\ 
0.20 & -2.8938 & -2.7436 & -2.7010 \\ 
0.30 & -3.1698 & -2.9990 & -2.9541 \\ 
0.40 & -3.4065 & -3.2211 & -3.1745 \\ 
0.50 & -3.6156 & -3.4184 & -3.3709 \\ 
0.60 & -3.8036 & -3.5969 & -3.5487 \\ 
0.70 & -3.9753 & -3.7606 & -3.7119 \\ 
0.80 & -4.1335 & -3.9122 & -3.8629 \\ 
0.90 & -4.2806 & -4.0538 & -4.0038 \\ 
1.00 & -4.4193 & -4.1908 & -4.1372 \\ 
2.00 & -5.4881 & -5.2306 & -5.1733 \\ 
3.00 & -6.2545 & -5.9827 & -5.9240 \\ 
5.00 & -7.3900 & -7.1035 & -7.0426 \\ 
7.00 & -8.2528 & -7.9582 & -7.8968 \\ 
10.00 & -9.2773 & -8.9754 & -8.9134 \\ 
20.00 & -11.6315 & -11.3192 & -11.2567 \\ 
30.00 & -13.2562 & -12.9403 & -12.8772 \\ 
50.00 & -15.5955 & -15.2756 & -15.2123 \\ 
70.00 & -17.3302 & -17.0081 & -16.9442 \\ 
100.00 & -19.3495 & -19.0247 & -18.9603 
\end{tabular}	
\end{table}
\begin{table}[t]
\caption{HF-ETG total energies for the first three excited states of
$(-2,+,-1)$ symmetry.
Zero field quantum numbers are given at the top of each column.
}\label{tab:he5}
\begin{tabular}{rrrr}
$\beta_Z$&${\rm 1s3d}_{-2}$&${\rm 1s4d}_{-2}$&${\rm 1s5d}_{-2}$\\
\hline
0.01 & -2.1659 & -2.0990 & -2.0742 \\ 
0.05 & -2.4320 & -2.2801 & -2.2356 \\ 
0.10 & -2.6871 & -2.4680 & -2.4120 \\ 
0.20 & -3.1005 & -2.7822 & -2.7141 \\ 
0.30 & -3.4394 & -3.0438 & -2.9685 \\ 
0.40 & -3.7309 & -3.2701 & -3.1902 \\ 
0.50 & -3.9890 & -3.4709 & -3.3873 \\ 
0.60 & -4.2215 & -3.6524 & -3.5657 \\ 
0.70 & -4.4342 & -3.8187 & -3.7288 \\ 
0.80 & -4.6306 & -3.9725 & -3.8814 \\ 
0.90 & -4.8134 & -4.1162 & -4.0229 \\ 
1.00 & -4.9866 & -4.2549 & -4.1558 \\ 
2.00 & -6.3268 & -5.3073 & -5.1947 \\ 
3.00 & -7.2952 & -6.0671 & -5.9467 \\ 
5.00 & -8.7385 & -7.1975 & -7.0674 \\ 
7.00 & -9.8406 & -8.0594 & -7.9229 \\ 
10.00 & -11.1540 & -9.0844 & -8.9411 \\ 
20.00 & -14.1875 & -11.4426 & -11.2871 \\ 
30.00 & -16.2905 & -13.0715 & -12.9091 \\ 
50.00 & -19.3286 & -15.4169 & -15.2460 \\ 
70.00 & -21.5954 & -17.1561 & -16.9805 \\ 
100.00 & -24.2283 & -19.1800 & -18.9982
\end{tabular}
\end{table}	
\begin{table}[t]
\caption{HF-ETG total energies for the first three excited states of
$(-2,-,-1)$ symmetry.
Zero field quantum numbers are given at the top of each column.
}\label{tab:he6}
\begin{tabular}{rrrr}
$\beta_Z$&${\rm 1s4f}_{-2}$&${\rm 1s5f}_{-2}$&${\rm 1s6f}_{-2}$\\
\hline
0.01 & -2.1201 & -2.0829 & -2.0636 \\ 
0.05 & -2.3268 & -2.2521 & -2.2241 \\ 
0.10 & -2.5299 & -2.4317 & -2.3985 \\ 
0.20 & -2.8637 & -2.7372 & -2.6990 \\ 
0.30 & -3.1378 & -2.9933 & -2.9522 \\ 
0.40 & -3.3738 & -3.2149 & -3.1726 \\ 
0.50 & -3.5823 & -3.4142 & -3.3690 \\ 
0.60 & -3.7702 & -3.5933 & -3.5469 \\ 
0.70 & -3.9417 & -3.7577 & -3.7100 \\ 
0.80 & -4.1000 & -3.9096 & -3.8610 \\ 
0.90 & -4.2473 & -4.0513 & -4.0020 \\ 
1.00 & -4.3861 & -4.1853 & -4.1353 \\ 
2.00 & -5.4573 & -5.2258 & -5.1717 \\ 
3.00 & -6.2263 & -5.9786 & -5.9225 \\ 
5.00 & -7.3654 & -7.0998 & -7.0416 \\ 
7.00 & -8.2310 & -7.9553 & -7.8957 \\ 
10.00 & -9.2549 & -8.9729 & -8.9126 \\ 
20.00 & -11.6182 & -11.3183 & -11.2562 \\ 
30.00 & -13.2459 & -12.9392 & -12.8766 \\ 
50.00 & -15.5879 & -15.2748 & -15.2119 \\ 
70.00 & -17.3240 & -17.0073 & -16.9441 \\ 
100.00 & -19.3448 & -19.0248 & -18.9605
\end{tabular}
\end{table}	
\begin{table}[t]
\caption{HF-ETG total energies for the first three excited states of
$(-3,+,-1)$ symmetry.
Zero field quantum numbers are given at the top of each column.
}\label{tab:he7}
\begin{tabular}{rrrr}
$\beta_Z$&${\rm 1s4f}_{-3}$&${\rm 1s5f}_{-3}$&${\rm 1s6f}_{-3}$\\
\hline
0.01 & -2.1406 & -2.0914 & -2.0651 \\ 
0.05 & -2.3923 & -2.2721 & -2.2320 \\ 
0.10 & -2.6344 & -2.4587 & -2.4085 \\ 
0.20 & -3.0275 & -2.7726 & -2.7110 \\ 
0.30 & -3.3511 & -3.0339 & -2.9655 \\ 
0.40 & -3.6294 & -3.2598 & -3.1872 \\ 
0.50 & -3.8759 & -3.4606 & -3.3846 \\ 
0.60 & -4.0980 & -3.6418 & -3.5630 \\ 
0.70 & -4.3019 & -3.8082 & -3.7267 \\ 
0.80 & -4.4906 & -3.9618 & -3.8784 \\ 
0.90 & -4.6669 & -4.1056 & -4.0200 \\ 
1.00 & -4.8318 & -4.2444 & -4.1532 \\ 
2.00 & -6.1195 & -5.2967 & -5.1919 \\ 
3.00 & -7.0517 & -6.0567 & -5.9443 \\ 
5.00 & -8.4426 & -7.1872 & -7.0654 \\ 
7.00 & -9.5058 & -8.0493 & -7.9220 \\ 
10.00 & -10.7734 & -9.0743 & -8.9396 \\ 
20.00 & -13.7044 & -11.4330 & -11.2861 \\ 
30.00 & -15.7378 & -13.0624 & -12.9073 \\ 
50.00 & -18.6784 & -15.4083 & -15.2443 \\ 
70.00 & -20.8674 & -17.1463 & -16.9778 \\ 
100.00 & -23.4342 & -19.1738 & -18.9986
\end{tabular}
\end{table}
\begin{table}[t]
\caption{HF-ETG total energies for the first three excited states of
$(-3,-,-1)$ symmetry.
Zero field quantum numbers are given at the top of each column.
}\label{tab:he8}
\begin{tabular}{rrrr}
$\beta_Z$&${\rm 1s5g}_{-3}$&${\rm 1s6g}_{-3}$&${\rm 1s7g}_{-3}$\\
\hline
0.01 & -2.1089 & -2.0783 & -2.0534 \\ 
0.05 & -2.3120 & -2.2476 & -2.2225 \\ 
0.10 & -2.5133 & -2.4275 & -2.3976 \\ 
0.20 & -2.8451 & -2.7332 & -2.6975 \\ 
0.30 & -3.1181 & -2.9895 & -2.9505 \\ 
0.40 & -3.3530 & -3.2120 & -3.1713 \\ 
0.50 & -3.5607 & -3.4103 & -3.3678 \\ 
0.60 & -3.7479 & -3.5897 & -3.5457 \\ 
0.70 & -3.9189 & -3.7539 & -3.7090 \\ 
0.80 & -4.0770 & -3.9060 & -3.8602 \\ 
0.90 & -4.2282 & -4.0478 & -4.0014 \\ 
1.00 & -4.3664 & -4.1818 & -4.1341 \\ 
2.00 & -5.4393 & -5.2229 & -5.1707 \\ 
3.00 & -6.2096 & -5.9759 & -5.9216 \\ 
5.00 & -7.3514 & -7.0976 & -7.0410 \\ 
7.00 & -8.2188 & -7.9533 & -7.8953 \\ 
10.00 & -9.2476 & -8.9714 & -8.9121 \\ 
20.00 & -11.6106 & -11.3173 & -11.2552 \\ 
30.00 & -13.2408 & -12.9382 & -12.8765 \\ 
50.00 & -15.5847 & -15.2745 & -15.2119 \\ 
70.00 & -17.3216 & -17.0066 & -16.9440 \\ 
100.00 & -19.3436 & -19.0249 & -18.9606
\end{tabular}
\end{table}

\begin{table}[t]
\caption{Comparison of Released--Phase QMC results, $E_{RP}$, with the current
ETG Hartree--Fock energies, $E_{HF}^{ETG}$, and the best previous RP
results from Ref. 18.  Blanks are left for the entries for which
no previous RP calculations were done.  Fixed--Phase QMC results, $E_{FP}$, are
also shown for the lowest excited state of each symmetry.
We note that the previous RP results, Ref. 18, are considerably
improved for the higher excitations, which reflect the higher
quality of the present HF wave-functions.}\label{tab:rp}
\begin{tabular}{lllllllllll}
&\multicolumn{4}{c}{${\rm 1s2s}$}&\multicolumn{3}{c}{${\rm 1s3s}$}
 &\multicolumn{3}{c}{${\rm 1s4s}$}\\
\cline{2-5}\cline{6-8}\cline{9-11}
\\
$\beta_Z$&$-E_{HF}^{ETG}$&Ref. 18&$-E_{RP}$&$-E_{FP}$&$-E_{HF}^{ETG}$&Ref. 18&$-E_{RP}$
 &$-E_{HF}^{ETG}$&Ref. 18&$-E_{RP}$\\
\tableline
0.01&2.2425&2.2438(3)&2.2439(2)&2.2440(2)
 &2.1199&2.1209(1)&2.1206(3)&2.0825&2.0687(9)&2.0830(6)\\
0.10&2.5720&2.5737(3)&2.5738(2)&2.5731(3)
 &2.4425&2.4395(9)&2.4433(3)&2.4034&2.3497(21)&2.4034(4)\\
1.00&4.3202&4.3204(5)&4.3218(6)&4.3217(3)
 &4.1708&4.1169(19)&4.1716(9)&4.1299&&4.1307(9) \\
\\
&\multicolumn{4}{c}{${\rm 1s2p}_0$}&\multicolumn{3}{c}{${\rm 1s3p}_0$}
 &\multicolumn{3}{c}{${\rm 1s4p}_0$}\\
\cline{2-5}\cline{6-8}\cline{9-11}
0.01&2.2031&2.2050(6)&2.2053(4)&2.2061(3)
 &2.1105&2.1100(3)&2.1116(2)&2.0724&2.0578(14)&2.0718(5) \\
0.10&2.6347&2.6397(9)&2.6395(5)&2.6385(3)
 &2.4537&2.4545(6)&2.4558(4)&2.4067&2.3925(7) &2.4077(4) \\
1.00&4.5283&4.5314(4)&4.5352(6)&4.5347(2)
 &4.2056&4.1651(8)&4.2067(7)&4.1417&      &4.1425(10) \\
\\
&\multicolumn{4}{c}{${\rm 1s2p}_{-1}$}&\multicolumn{3}{c}{${\rm 1s3p}_{-1}$}
 &\multicolumn{3}{c}{${\rm 1s4p}_{-1}$}\\
\cline{2-5}\cline{6-8}\cline{9-11}
0.01&2.2353&2.2380(9)&2.2384(3)&2.2380(2)
 &2.1220&2.1219(5)&2.1231(5)&2.0888&2.0809(9)&2.0898(3) \\
0.10&2.8295&2.8354(5)&2.8354(6)&2.8356(2)
 &2.4855&2.4856(5)&2.4869(11)&2.4178&2.3852(17)&2.4189(2) \\
1.00&5.4000&5.4072(13)&5.4101(22)&5.4096(3)
 &4.2742&4.2436(32)&4.2753(2)&4.1605&--       &4.1608(8) \\
\\
&\multicolumn{4}{c}{${\rm 1s3d}_{-1}$}&\multicolumn{3}{c}{${\rm 1s4d}_{-1}$}
 &\multicolumn{3}{c}{${\rm 1s5d}_{-1}$}\\
\cline{2-5}\cline{6-8}\cline{9-11}
0.01&2.1403&2.1410(1)&2.1412(3)&2.1414(3)
 &2.0903&2.0912(2)&2.0912(2)&2.0673&2.0624(9)&2.0696(4) \\
0.10&2.5528&2.5590(27)&2.5583(18)&2.5577(4)
 &2.4382&2.4369(12)&2.4392(5)&2.4012&2.3504(14)&2.4022(4) \\
1.00&4.4193&4.4214(19)&4.4236(8)&4.4235(5)
 &4.1908&4.1074(9)&4.1919(6)&4.1372&--       &4.1379(7) \\
\\
&\multicolumn{4}{c}{${\rm 1s3d}_{-2}$}&\multicolumn{3}{c}{${\rm 1s4d}_{-2}$}
 &\multicolumn{3}{c}{${\rm 1s5d}_{-2}$}\\
\cline{2-5}\cline{6-8}\cline{9-11}
0.01&2.1659&&2.1667(2)&2.1664(2) 
 &2.0990&& 2.0996(2)&2.0742&&2.0751(4) \\
0.10&2.6871&&2.6899(9)&2.6902(2) 
 &2.4680&& 2.4691(14)&2.4120&&2.4125(4) \\
1.00&4.9866&&4.9941(9)&4.9956(5)
 &4.2549&& 4.2565(21)&4.1558&&4.1567(5) \\
\\
&\multicolumn{4}{c}{${\rm 1s4f}_{-2}$}&\multicolumn{3}{c}{${\rm 1s5f}_{-2}$}
 &\multicolumn{3}{c}{${\rm 1s6f}_{-2}$}\\
\cline{2-5}\cline{6-8}\cline{9-11}
0.01&2.1201&&2.1198(8)&2.1206(1) 
 &2.0829&&2.0834(5)&2.0636&&2.0655(6) \\
0.10&2.5299&&2.5316(9)&2.5311(3) 
 &2.4317&&2.4326(3)&2.3985&&2.3993(4) \\
1.00&4.3861&&4.3888(22)&4.3907(5)
 &4.1853&&4.1854(3)&4.1353&&4.1356(1) \\
\\
&\multicolumn{4}{c}{${\rm 1s4f}_{-3}$}&\multicolumn{3}{c}{${\rm 1s5f}_{-3}$}
 &\multicolumn{3}{c}{${\rm 1s6f}_{-3}$}\\
\cline{2-5}\cline{6-8}\cline{9-11}
0.01&2.1406&&2.1421(4)&2.1422(3) 
 &2.0914&&2.0920(7)&2.0651&&2.0680(7) \\
0.10&2.6344&&2.6330(22)&2.6361(4) 
 &2.4587&&2.4591(14)&2.4085&&2.3980(8) \\
1.00&4.8318&&4.8392(13)&4.8407(4) 
 &4.2444&&4.2459(9)&4.1532&&4.1537(2) \\
\\
&\multicolumn{4}{c}{${\rm 1s5g}_{-3}$}&\multicolumn{3}{c}{${\rm 1s6g}_{-3}$}
 &\multicolumn{3}{c}{${\rm 1s7g}_{-3}$}\\
\cline{2-5}\cline{6-8}\cline{9-11}
0.01&2.1089&&2.1095(3)&2.1094(2) 
 &2.0783&&2.0791(3)&2.0534&&2.0544(12)\\
0.10&2.5133&&2.5141(2)&2.5141(1)  
 &2.4275&&2.4281(2)&2.3976&&2.3982(8) \\
1.00&4.3664&&4.3714(18)&4.3694(4) 
 &4.1818&&4.1835(6)&4.1341&&4.1349(6) \\
\end{tabular}
\end{table}
%
\begin{table}
\caption{Comparison of fully correlated RPQMC energies, $E_{RP}$, with recent
values found in the literature, where available.
Values in parentheses for the RPQMC energies are statistical error bars.
Those for Ref. [19] are estimated uncertainties in the last digit quoted.
Zero field quantum numbers are listed above each block of entries.
All energies are in hartree atomic units.}\label{tab:rp_comp}
\begin{tabular}{llllllllll}
$\beta_Z$&$-E_{RP}$&Ref. [19]&Ref. [20]&$-E_{RP}$&Ref. [19]&Ref. [20]
 &$-E_{RP}$&Ref. [19]&Ref. [20]\\
&\multicolumn{3}{c}{${\rm 1s2s}$}&\multicolumn{3}{c}{${\rm 1s3s}$}
 &\multicolumn{3}{c}{${\rm 1s4s}$}\\
\cline{2-4}\cline{5-7}\cline{8-10}
0.01&2.2439(2)&&2.243958&2.1206(3)&&2.121107&2.0830(6)&&2.087409 \\
0.10&2.5738(2)&2.57859(1)&2.573615&2.4433(3)&2.4686(56)&2.443352&2.4034(4)&&2.403631 \\
&\multicolumn{3}{c}{${\rm 1s2p}_0$}&\multicolumn{3}{c}{${\rm 1s3p}_0$}
 &\multicolumn{3}{c}{${\rm 1s4p}_0$}\\
\cline{2-4}\cline{5-7}\cline{8-10}
0.01&2.2053(4)&&2.205130&2.1116(2)&&2.111478&2.0718(5)&&2.079242  \\
0.10&2.6395(5)&2.64014(0)&2.638222&2.4558(4)&2.4944(90)&2.455054&2.4077(4)&2.411(28)&2.407425  \\
&\multicolumn{3}{c}{${\rm 1s2p}_{-1}$}&\multicolumn{3}{c}{${\rm 1s3p}_{-1}$}
 &\multicolumn{3}{c}{${\rm 1s4p}_{-1}$}\\
\cline{2-4}\cline{5-7}\cline{8-10}
0.01&2.2384(3)& & &2.1231(5)& & &2.0898(3)  \\
0.10&2.8354(6)&2.83572(0)& &2.4869(11)&2.5044(3)& &2.4189(2)&2.396(45)  \\
&\multicolumn{3}{c}{${\rm 1s3d}_{-1}$}&\multicolumn{3}{c}{${\rm 1s4d}_{-1}$}
 &\multicolumn{3}{c}{${\rm 1s5d}_{-1}$}\\
\cline{2-4}\cline{5-7}\cline{8-10}
0.01&2.1412(3)& & &2.0912(2)& & &2.0696(4)  \\
0.10&2.5583(18)&2.56570(7)&&2.4392(5)&2.437(14)& &2.4022(4)  
\end{tabular}
\end{table}
\end{document}